\documentclass[journal, 10pt]{IEEEtran}
\usepackage{graphicx} 
\usepackage[utf8]{inputenc} %
\usepackage[T1]{fontenc}    %
\usepackage{hyperref}       %
\usepackage{url}            %
\usepackage{booktabs}       %
\usepackage{amsfonts}       %
\usepackage{nicefrac}       %
\usepackage{microtype}      %
\usepackage{caption}
\usepackage{microtype}
\usepackage{graphicx}
\usepackage{subcaption}
\usepackage{booktabs} %
\usepackage{enumitem} 
\usepackage{bm}
\usepackage{placeins}
\usepackage{algorithmicx,algpseudocode,algorithm}
\usepackage{wrapfig,tikz}
\usepackage{amsmath,longtable,fancyhdr,booktabs,multirow,graphicx,float}
\usepackage{adjustbox, bigstrut, tabularx, multirow, makecell, diagbox}
\usepackage{amssymb,xcolor,amsthm}
\usepackage{color}
\usepackage{colortbl}
\usepackage{theoremref}
\usepackage{stmaryrd}
\usepackage{xcolor}
\usepackage{amsthm}

\usepackage{cuted}
\usepackage{balance}

\ifCLASSINFOpdf
\else
\fi
\onecolumn

\begin{document}
%
\title{A Mixture-of-Experts Framework for Multi-Stream Quickest Change Detection}
%
%
%

\author{Wuxia~Chen, Arman~Adibi, Taposh~Banerjee, and Vahid~Tarokh

\thanks{Wuxia Chen and Taposh Banerjee are with the Department of Industrial Engineering, University of Pittsburgh, Pittsburgh, PA 15260 (E-mail: WUC3@pitt.edu, taposh.banerjee@pitt.edu). }
\thanks{Arman Adibi is with the School of Computer and Cyber Sciences at Augusta University, Augusta University, Augusta, GA 30901 (E-mail: aadibi@augusta.edu).}
\thanks{Vahid Tarokh is with the Department of Electrical and Computer Engineering, Duke University, Durham, NC 27708 (Email: vahid.tarokh@duke.edu).}
}

\maketitle

\begin{abstract}
A Mixture-of-Experts (MoE) framework is proposed for quickest change detection (QCD) where there are multiple experts, and each expert produces a sequence of statistics for change detection. A change is declared when a weighted sum of these statistics crosses a threshold. This framework is then applied to the problem of multi-stream QCD
to obtain a family of powerful algorithms that outperform the best known algorithms in the literature. 
In the multi-stream QCD problem, a decision-maker observes multiple streams of observations. At the change point, the distribution of an unknown subset of the streams changes. The optimal algorithm based on a generalized likelihood ratio is computationally expensive.  One of the most computationally and performance-efficient algorithms in the literature is the rank-weighted sum (RWS) scheme based on Shiryaev-Roberts statistics. 
It is shown in the paper that the MoE-based QCD algorithm for multi-stream data outperforms the RWS scheme. This is demonstrated through simulation results on real-valued Gaussian mean shift data and high-dimensional data from Gaussian--Bernoulli Restricted Boltzmann Machines (GB-RBMs).

%

\end{abstract}


%
\IEEEpeerreviewmaketitle

\section{Introduction}


Detecting abrupt changes in sequential data is a fundamental problem with numerous applications, including power system monitoring, quality control, target identification, biology, cyber-security, and finance\cite{chen2015quickest, hawkins2003changepoint, blackman2004multiple, siegmund2013change}. In statistics, this problem is formulated as the problem of quickest change detection (QCD), where the objective is to detect a change in the underlying distribution of a stochastic process as quickly as possible while controlling the frequency of false alarms \cite{shiryaev1963optimum}.

One of the most prominent algorithms for QCD is the Cumulative Sum (CUSUM) procedure \cite{page1954continuous}, which is particularly effective when the observations are independent and identically distributed (i.i.d.) before and after the change. Owing to its recursive structure, the CUSUM algorithm is computationally efficient and enjoys strong theoretical guarantees. In particular, it is optimal in minimizing the worst-case detection delay under Lorden's criterion while satisfying a prescribed false-alarm constraint \cite{lorden1971procedures,moustakides1986optimal}. Another popular algorithm is the Shiryaev-Roberts algorithm, which is optimal or asymptotically optimal under generalized Bayes and minimax criteria \cite{pollak1985optimal}. Comprehensive surveys of QCD and related optimal procedures can be found in \cite{veeravalli2014quickest, tartakovsky2014sequential, tartakovsky2019sequential, poor2008quickest}.

In many modern applications, observations are collected simultaneously from multiple sensors or data streams. Examples include sensor networks, communication systems, industrial monitoring systems, financial markets, and video surveillance \cite{xie2021sequential}. In such settings, a change may affect only an unknown subset of the streams, while the remaining streams continue to follow their original distributions. An asymptotically optimal test for this problem is based on the generalized likelihood ratio (GLR) statistic \cite{fellouris2016second}. In the GLR test, a CUSUM statistic is computed for each possible post-change subset, and the change is declared when any one of the CUSUM statistics crosses a pre-defined threshold. Since the post-change subset can be arbitrary, the potential number of CUSUM tests that must be run in parallel can be exponential in the number of streams. Thus, this test, in general, is computationally expensive. Therefore, a central problem in multi-stream QCD is how to obtain computationally efficient algorithms that match the performance of the GLR test. 

A common approach taken in the past several years has been to construct a local monitoring statistic for each stream and then combine or fuse them to obtain a global statistic. 
Some important and well-studied rules are MAX \cite{tartakovsky2006detection}, SUM \cite{mei2010efficient}, TOP-r \cite{mei2011quickest}, and order-filtering procedures \cite{li2020efficient}.
More recently, rank-weighted fusion rules have been proposed for local Shiryaev--Roberts--Pollak (SRP) statistics \cite{ZhangMei-SQA2026}. In this approach, the local SRP statistics are ordered from largest to smallest at each time point. The $k$-th ordered statistic is multiplied by its rank $k$, and the resulting weighted statistics are summed to form a global monitoring statistic. This linear rank-weighted rule is simple, tuning-free, and has demonstrated strong empirical performance. However, the weight function used in \cite{ZhangMei-SQA2026} based on linear ranks, represents only one particular choice among many possible alternatives. 

This observation motivates a Mixture-Of-Experts (MoE) architecture for multi-stream QCD, in which there are several experts; each expert produces a monitoring statistic, and a fusion rule or mixture is used to obtain a global statistic.



In this work, we first propose a MoE architecture for the general QCD problem, not necessarily for multi-stream QCD. We believe that this architecture can be used to find powerful algorithms applicable to a wide range of applications. The proposed QCD-MoE architecture naturally raises the question of how the experts and their weights should be selected, with the possibility of the weights being adapted online based on the past observed data. However, the focus of this work is merely to establish the effectiveness of the MoE framework. Thus, the issues of expert selection and weight optimization will be addressed in a future report.

To show the effectiveness of the proposed MoE architecture, we report a case study of the multi-stream QCD problem. Specifically, by treating each stream statistic as an expert, we optimize over the choice of the mixture coefficients. The expert statistic used for each stream is the Shiryaev-Roberts statistic, as in \cite{ZhangMei-SQA2026}. We show that this optimization over the mixture coefficients leads to better performance as compared to the rank-weighted Shiryaev-Roberts scheme studied in \cite{ZhangMei-SQA2026}. 
Through extensive numerical experiments, we demonstrate that the linear rank-weighted rule of \cite{ZhangMei-SQA2026} is not uniformly optimal, and one can get improved performance by using nonlinear weights. These findings provide further motivation for exploring the general MoE framework for QCD. 

To summarize, the main contributions of this paper are as follows:
\begin{itemize}
\item We propose a general mixture-of-experts (MoE) framework for QCD,  in which each expert produces a monitoring statistic and the final decision is based on a mixture of these expert statistics.

\item We apply the general framework to the multi-stream QCD problem by constructing a family of MoE-based QCD algorithms using Shiryaev--Roberts statistics. Each MoE-based algorithm uses distinct mixture weights, including logarithmic, linear, and polynomial weights.

\item Through extensive numerical experiments, we demonstrate that no fixed mixture choice is uniformly optimal across all sparsity regimes. Different MoE-based algorithms achieve the best detection performance under different change scenarios. However, the best MoE-based algorithm can substantially outperform the conventional linear rank-weighted rule from \cite{ZhangMei-SQA2026} under the same false-alarm constraint.

\item We further apply the proposed MoE framework to score-based statistics \cite{Banerjee_SQA_2024}, showing that the same qualitative behavior is observed beyond likelihood-ratio-based statistics.

\end{itemize}

The remainder of the paper is organized as follows. Section~\ref{sec:general_MoE} introduces a general mixture-of-experts (MoE) fusion architecture for quickest change detection. Section~\ref{sec:application_multi_qcd} applies this framework to multi-stream quickest change detection by constructing rank-weighted experts based on local Shiryaev--Roberts statistics. Section~\ref{sec:numerical_results} presents numerical comparisons of different rank-weighting algorithms and investigates their detection performance under various change scenarios. Section~\ref{sec:score-based Fusion} extends the same framework to score-based local detection statistics and presents the corresponding numerical results. Finally, Section~\ref{sec:conclusion} concludes the paper and discusses future research directions.

\section{General Mixture-Of-Experts Architecture for Quickest Change Detection}
\label{sec:general_MoE}

In this section, we introduce a general mixture-of-experts (MoE) architecture for quickest change detection (QCD). In this framework, there are multiple experts, each with its own decision statistics. A weighted sum or a mixture of these statistics is used for QCD.  
This framework is sufficiently general to accommodate different choices of experts and the corresponding weights.  In Section~\ref{sec:application_multi_qcd}, we will apply this framework to multi-stream QCD to obtain efficient algorithms. 

%
%
%

In the MoE framework, there are $L$ experts. The expert $e$ with $e\in\{1,\ldots,L\}$ produces at time $n$ a statistic 
\[
G_n^{(e)}.
\]
The statistics $\{G_n^{(e)}\}_{n\geq 1}$ need not have a physical interpretation and do not have to be change-detection statistics by themselves. 
At each time $n$, the $L$ experts produce $L$ different sequences of statistics
\[
\{G_n^{(1)}\}_{n\geq 1}, \{G_n^{(2)}\}_{n\geq 1}, \dots, \{G_n^{(L)}\}_{n\geq 1}.
\]
The contribution from an expert $e$ at time $n$ is 
$$
\pi_n^{(e)} G_n^{(e)},
$$
where $\pi_n^{(e)}$ is a an appropriate time and data-dependent weight. 
In the MoE framework for QCD, these contributions from the $L$ experts are then combined to form a QCD statistic:
\begin{equation}
 \widetilde{G}_n = \sum_{e=1}^{L} \pi_n^{(e)}G_n^{(e)}.
 \label{eq:general_expert}
\end{equation}
The MoE-QCD stopping rule is
\[
T = \inf \{  n \ge 1:  \widetilde{G}_n \ge A \},
\]
where $A$ is a prescribed detection threshold to control the rate of false alarms. In a slightly more general setting, the threshold $A$ can also be made data- and time-dependent. Thus, the objective of the MoE QCD framework is to combine the information provided by the individual experts into a single global decision statistic. Rather than relying on a single monitoring procedure throughout the detection process, the framework allows multiple experts to contribute simultaneously to the final decision.

To motivate this architecture, consider the case where the statistics $\{G_n^{(e)}\}$ calculated by the expert $e$ do directly capture the evidence of a change. Then the statistics can be directly compared with a threshold to declare the change. In other words, each expert can then be seen as an algorithm in itself. We consider some scenarios:
\begin{itemize}
    \item In multi-stream QCD, different experts may correspond to different fusion strategies. One expert may be more effective when changes are sparse, while another may perform better when a larger number of components are affected. Since the underlying change scenario is typically unknown in advance, relying on a single expert may lead to suboptimal detection performance. However, adaptively adjusting the mixture coefficients based on data in real time might help us circumvent this issue. 
    \item In parametric and non-parametric QCD, where the precise modeling information is unknown, different experts may correspond to different modeling assumptions. Since the true model is generally unknown, one can use a mixture of statistics generated, one for each model, to improve its chances of consistent change detection. 
    \item In QCD, there are various problem formulations. Each formulation leads to its own optimal algorithm. To a practitioner, it is not always clear which criterion is more appropriate, e.g., Lorden's minimax criterion \cite{lorden1971procedures}, Pollak's minimax criterion \cite{pollak1985optimal}, or a generalized-Bayes formulation \cite{pollak2009optimality}. With several optimal algorithms to choose from, the practitioner might prefer using a mixture of the statistics from these algorithms (appropriately transformed) to detect the change, and optimize over the mixture coefficients in real time.  
\end{itemize}
 
In the distributed QCD setting \cite{veeravalli2001decentralized, mei2005information, mei2011quickest, tartakovsky2008asymptotically, banerjee2015data}, different local decision-makers may be viewed as different experts.

In this paper, we do not discuss how to construct experts in general, nor do we suggest concrete algorithms for updating the mixture coefficient in an online fashion. However, we do suggest the following:
\begin{enumerate}
    \item The mixture coefficients be updated over multiple end-to-end simulations of the change detection process. 
    \item The optimal mixture coefficients will be a time-varying non-linear function of all the past history of the data collected. One way to approximate such a function would be to use a parametric family or an artificial neural network.  
\end{enumerate}

Overall, our focus in this paper is to show that the MoE architecture is promising by reporting some positive results in the multi-stream QCD problem.





\section{Application of the MoE Framework to Multi-Stream QCD}
\label{sec:application_multi_qcd}
In this section, we apply the general mixture-of-experts (MoE) framework introduced in Section~\ref{sec:general_MoE} to the multi-stream quickest change detection (QCD) problem. We first formulate the multi-stream change-point model and review the local Shiryaev--Roberts (SRP) statistics. We then construct a family of rank-weighted experts based on different rank-weighting functions.

\subsection{Multi-Stream QCD and the Generalized Likelihood Ratio (GLR) Statistic}

Consider a system consisting of $K$ parallel data streams. At each time $n \geq 1$, the observation from stream $k$ is denoted by $X_{k,n}$, where $k \in \{1,\ldots,K\}$.
Before an unknown change point $\nu$, all observations follow the pre-change density $f$. At time $\nu$, an unknown subset $\mathcal{A} \subseteq \{1,\ldots,K\}$ of streams undergoes a distributional change from $f$ to a post-change density $g$. Thus,

\[
X_{k,n} \sim
\begin{cases}
f, & n < \nu,\\
g, & n \geq \nu,\ k \in \mathcal{A},\\
f, & n \geq \nu,\ k \notin \mathcal{A}.
\end{cases}
\]
Conditioned on the change point, the observations are assumed to be independent across time and across streams.
The notation $\mathbb{P}_{\nu}$ and $\mathbb{E}_{\nu}$ are used to denote the probability measure and expectation when the change occurs at time $\nu$. Under the no-change measure, denoted by $\mathbb{P}_{\infty}$, all observations follow the pre-change density $f$, and the corresponding expectation is denoted by $\mathbb{E}_{\infty}$.

A detection procedure is a stopping time $T$ adapted to the filtration generated by the observations. In the formulation of Lorden \cite{lorden1971procedures}, the false-alarm performance of $T$ is measured by the average run length (ARL) to false alarm,
\[
\text{ARL}(T) :=\mathbb{E}_{\infty}[T].
\]
The delay of the stopping time is measured by the Worst-case Average Detection Delay (WADD) metric:
$$
\text{WADD}(T) = \sup_\nu \text{ess sup} \;  \mathbb{E}_\nu [(T - \nu +1)^+ | \mathcal{F}_{\nu-1}],
$$
where $\mathcal{F}_{\nu-1}$ is the sigma-algebra generated by data collected up to time $\nu-1$ and the essential supremum of a random variable $Y$, $\text{ess sup } Y$, is the smallest $C$ such that $\mathbb{P}(Y \leq C)=1$. 
The goal of the multi-stream QCD problem is to solve
\begin{equation}
    \begin{split}
        \min_T \; \text{WADD}(T), \quad \text{subj. to} \; \text{ARL}(T) \geq \gamma.
    \end{split}
\end{equation}
Here $\gamma$ is a user-specified constraint on the ARL.

A rule that is known to be asymptotically optimal for this problem is the GLR rule given by \cite{fellouris2016second, tartakovsky2014sequential, tartakovsky2019sequential}
\begin{equation}
\label{eq:GLR}
    \begin{split}
        T_\text{GLR} = \min\left\{n \geq 1: \max_{\mathcal{A}} \max_{1 \leq k \leq n} \sum_{\ell \in \mathcal{A}} \sum_{i=k}^n \log \frac{g(X_{i,\ell})}{f(X_{i,\ell})} \geq A\right\}.
    \end{split}
\end{equation}
However, since the subset $\mathcal{A}$ is arbitrary, the external maximum over $\mathcal{A}$ involves searching over $2^K$ possibilities. 
This complexity can be brought down to polynomial in $K$ if we assume that $|\mathcal{A}| \leq m \ll K$.

\subsection{Linear Rank-Weighted Shiryaev-Roberts Procedure}
The computational complexity of the generalized likelihood ratio procedure \eqref{eq:GLR} has motivated several works with lower complexity. Prominent among these are the SUM procedure in \cite{mei2010efficient} and the 
Linear Rank-Weighted Sum (RWS) Shiryaev-Roberts Procedure in \cite{ZhangMei-SQA2026}. We discuss the latter in this section, since according to \cite{ZhangMei-SQA2026}, the RWS scheme generally outperforms the SUM scheme and most other multi-stream QCD algorithms. 

In the RWS procedure, for each stream $k = 1,2, \dots K$, the Shiryaev--Roberts statistic is initialized at $R_{k,0}=0$ and updated recursively as
\[
R_{k,n} = (R_{k,n-1}+1)  \frac{g(X_{k,n})}{f(X_{k,n})}, 
\]
Equivalently,
\[
R_{k,n} = \sum_{i=1}^{n} \prod_{j=i}^{n} \frac{g(X_{k,j})}{f(X_{k,j})}.
\]
At each time $n$, let
\[
R_{(1),n}
\geq
R_{(2),n}
\geq
\cdots
\geq
R_{(K),n}
\]
denote the ordered local statistics, where $R_{(1),n}$ is the largest and $R_{(K),n}$ is the smallest. The RWS stopping rule is
\begin{equation}
    \begin{split}
        T_{RWS} = \min\left\{n \geq 1: \sum_{k=1}^K k R_{(k),n} \geq A\right\}.
    \end{split}
\end{equation}

\subsection{The Proposed MoE-Based QCD Algorithm for Multi-Stream Data}
In this paper, we consider a MoE-based approach to this problem. We think of each stream's Shiryaev-Roberts statistics as the statistics from an expert, and consider the mixture
\begin{equation}
 G_n^{(w)}\triangleq \sum_{k=1}^{K} w(k)R_{(k),n}.
 \label{eq:fusion_rule}
\end{equation}
The corresponding stopping rule is
\begin{equation}
   T^{(w)} \triangleq \inf \{ n\ge 1: G_n^{(w)}\ge A_w \},
   \label{eq:stopping}
\end{equation}
where the threshold $A_w$ is selected to satisfy a prescribed average run length to false alarm constraint,
\[
\mathbb{E}_{\infty}\left[T^{(w)}\right]\ge \gamma.
\]
This generalizes the RWS scheme to allow for non-linear weights. That is, if we set $w(k)=k$, then we recover the RWS scheme from \cite{ZhangMei-SQA2026}. 
We remark that the weight choices here, and even in the RWS scheme, are dependent on the entire data history since the ranking is time-varying. 
Our approach is to study the impact of different choices of the mixture coefficients or the rank-weighting function $\{w(k)\}$ on the performance of this algorithm. For simplicity, we restrict our search to linear, logarithmic, and polynomial weights.



\clearpage
\section{Numerical Results for MoE-Based Multi-Stream QCD Algorithm}
\label{sec:numerical_results}


In this section, we evaluate the detection performance of the MoE-based multi-stream QCD algorithm proposed in \eqref{eq:stopping}. We first describe the simulation setup. We then compare the MoE-based multi-stream QCD algorithm with the RWS scheme from \cite{ZhangMei-SQA2026}. 
\begin{table*}[h!]
\centering
\caption{Effect of the fusion parameter $m$, for different weights $w(k)$, with $K=100$ and $\gamma=1000$.}
\label{tab:m_sensitivity}
\begin{tabular}{lcccccccccccc}
\hline
$m$ 
& 3 & 5 & 8 & 10 & 20& 30 & 40 & 50 & 60 & 70  & 80 & 100 \\
\hline
log (1+k)
& 47.18  & 40.74  & 35.90 & 33.90  & 28.35 & 25.32 & 23. 39 &  21.89  & 20.72 & 19.73 &18.97 & 17.64 \\ 
$\sqrt{k}$
& 47.66 & 41.18 & 36.27 & 34.10 & 28.52  & 25.55 &  23.51 & 21.96  &  20.75 & 19.80 &  18.94 & 17.50 \\ 
$k $
& 47.36 & 40.06 & 34.74 & 32.30 & 25.42 & 21.73 & 19.13 & 17.22 & 15.71 & 14.55  & 13.52 & 11.95 \\ 

$k^2$ 
& 59.80 &  50.42 & 41.72 & 38.21 & 26.11  & 20.07 &  16.49 & 14.03  & 12.31 & 10.92 &9.91  & 8.37 \\

$k^3$ 
&83.65  & 71.03   &59.25 & 53.47 &33.78 & 23.75&18.21  &14.90  &12.56 &  10.95&  9.7 &7.96  \\

$k^4$ 
& 107.98 &  92.70 & 78.23 & 70.88 & 42.89 &  27.99 & 20.36  &16.02 & 13.11 & 11.24 & 9.74 & 7.78 \\
\hline

\end{tabular}
\end{table*}

\begin{table*}[h!]
\centering
\caption{Best-performing rank-weighting functions or mixture for different numbers of affected streams for  $K=100$ and $\gamma=1000$}
\begin{tabular}{lcccccccccccc}
\hline
$m$ 
& 3 & 5 & 8 & 10 & 20 & 30 & 40 & 50 & 60 & 70& 80 & 100 \\
\hline
$k$ 
& 47.36 & 40.06 & 34.74 & 32.30 & 25.42  & 21.73  & 19.13 & 17.22 & 15.71 & 14.55 & 13.52 & 11.95 \\ 
\hline
Best delay
& 47.18   & 40.06  & 34.74 & 32.30  & 25.42  & 20.07 &  16.49  &  14.03 & 12.31 &  10.92&  9.7 & 7.78 \\ 

Corresponding mixture weights& $\log (1+k)$  & $k$  & $k$ & $k$  & $k$  & $k^2$ & $k^2$  &  $k^2$  & $k^2$&   $k^2$ &  $k^3$ & $k^4 $\\ 

\hline
\end{tabular}
\label{tab:best_performance}
\end{table*}

\subsection{Simulation Setup}
We consider a system with $K=100$ independent data streams. Before the change, the observations are independently generated according to
\[
X_{k,n}\sim \mathcal{N}(0,1).
\]
At the change point, the distributions of $m$ streams change to
\[
X_{k,n}\sim \mathcal{N}(0.5,1),
\]
while the remaining $K-m$ streams continue to follow the pre-change distribution. The identities of the affected streams are unknown. 
For each choice of mixture coefficients or mixture functions $\{w(k)\}$, the threshold $A_w$ in \eqref{eq:stopping} is chosen to satisfy the same average run length to false alarm constraint of $\gamma=1000$. 
All reported detection delays are estimated using 2500 independent Monte Carlo replications.

\subsection{Comparison of MoE QCD Algorithm with the RWS Scheme}
We investigate the following family of mixture coefficients $\{w(k)\}$ in \eqref{eq:stopping}:
\[
\log(1+k), 
\sqrt{k}, 
k, 
k^2,
k^3,
k^4.
\]
As discussed earlier, these choices lead to random data-dependent mixture coefficients, as the ranks of the Shiryaev-Roberts statistics are time-varying. 
Table~\ref{tab:m_sensitivity} reports the average detection delays for different numbers of affected streams.

The results show that the best choice of mixture coefficients (among the categories chosen) depend with the number of affected streams. When $m=3$, the logarithmic weight $\log(1+k)$ gives the smallest observed delay. For $m=5,8,10$, and $20$, the linear weight $w(k)=k$ gives the smallest delay. When the number of affected streams increases, higher-order polynomial weights become more effective. In particular, $w(k)=k^2$ gives the smallest delay for $m=30,\ldots,70$, while $w(k)=k^3$ and $w(k)=k^4$ give the smallest delays for $m=80$ and $m=100$, respectively.
Thus, no single choice of mixture weights performs best across all values of $m$. Although the linear weight $w(k)=k$ is effective when the number of affected streams is relatively small, a different mixture can achieve strictly shorter detection delays when more streams are affected. This provides direct evidence that the choice of rank-weighting function or the mixture is an important component of the fusion rule and that restricting the procedure to the linear weight may overlook substantial performance improvements. In other words, using a MoE-based QCD algorithm can give a significant performance boost.

\subsection{Best-Performing Mixture Weight Across Different Numbers of Affected Streams}

We next investigate how the best-performing mixture changes with the number of affected streams. We fix the total number of streams at $K=100$ and the average run length to false alarm at $\gamma=1000$, while varying the number of affected streams $m$.

For each value of $m$, Table~\ref{tab:best_performance} reports the detection delay obtained by the conventional linear weight $w(k)=k$, and the smallest detection delay achieved among all candidate mixture functions. The corresponding best-performing mixture function is also reported.

The results show that the best-performing mixture function depends strongly on $m$. When the number of affected streams is small, the linear weighting function performs well. For example, at $m=5,8,10$, and $20$, the smallest detection delay among the considered polynomial weights is attained by $w(k)=k$. In contrast, when more streams are affected, higher-order polynomial weights become increasingly effective. The quadratic weight $w(k)=k^2$ gives the smallest delay for $m=30,40,50,60,70$, while $k^3$ and $k^4$ achieve the smallest delays at $m=80$ and $m=100$, respectively.




\subsection{Performance Comparisons When All the Streams Are Affected}
In Table~\ref{tab:m_sensitivity} and Table~\ref{tab:best_performance}, we saw that the MoE-based technique provides the most benefit when the number of affected streams is large. In this subsection, we provide comparisons by changing the total number of streams and then causing a change in all of them.

\begin{table*}[h!]
\centering
\caption{Performance comparison  of rank-weighting functions for $m=10$, $K=10$, and $\gamma = 1000$.}
\label{tab:all_affected_10}
\begin{tabular}{lccccccccc}
\hline
 & $\log(1+k)$ & $\sqrt{k}$ & $k$ & $k^2$ & $k^3$ & $k^4$ & $k^5$ & $k^6$ & $e^{k}$ \\
\hline
Threshold $A_w$ 
& 5.52e+03 & 7.85e+03 & 8.45e+03 & 1.37e+04 & 5.69e+04 & 3.83e+05 & 2.98e+06 & 2.50e+07 & 4.12e+05
\\

$\mathbb{E}_\nu[T]$ 
&  21.92& 22.38 & 20.60 & 15.49 &13.23  & 13.06  & 13.12 & 13.16  & 13.66 \\
\hline
\end{tabular}
\end{table*}

\begin{table*}[h!]
\centering
\caption{Performance comparison of rank-weighting functions for $m=20$, $K=20$, and $\gamma = 1000$.}
\label{tab:all_affected_20}
\begin{tabular}{lccccccccc}
\hline
 & $\log(1+k)$ & $\sqrt{k}$ & $k$ & $k^2$ & $k^3$ & $k^4$ & $k^5$ & $k^6$ & $e^{k}$ \\
\hline
Threshold $A_w$ 
&1.13e+04 & 1.61e+04 & 1.85e+04 & 5.19e+04 & 5.16e+05 & 7.11e+06 & 1.09e+08 & 1.77e+09 & 4.82e+09
  \\

$\mathbb{E}_\nu[T]$ 
&  20.44& 20.67 & 17.72 & 11.82 & 10.73  & 10.66  & 10.63 & 10.75  & 11.88 \\
\hline
\end{tabular}
\end{table*}




Table~\ref{tab:all_affected_10} reports the results for $m=K=10$. The linear weighting function $w(k)=k$ yields an average detection delay of $20.60$. In comparison, the polynomial weights $k^2$, $k^3$, and $k^4$ reduce the average detection delay to $15.49$, $13.23$, and $13.06$, respectively. Increasing the polynomial order beyond four provides little additional improvement in this setting.

Table~\ref{tab:all_affected_20} reports the corresponding results for $m=K=20$. The linear weight gives an average detection delay of $17.72$, whereas the polynomial weights $k^2$, $k^3$, $k^4$, and $k^5$ reduce the delay to $11.82$, $10.73$, $10.66$, and $10.63$, respectively.


\section{Numerical Results for Unnormalized and Score-Based Models}
\label{sec:score-based Fusion}
In this section, we compare the performance of the MoE-based QCD algorithm with the RWS scheme for non-Gaussian high-dimensional data. We show that the MoE-based scheme shows similar improvements as we saw in the case of one-dimensional Gaussian mean shift data. 

\begin{table*}[h!]
\centering
\caption{Effect of the fusion parameter $m$, for different weights $w(k)$, with $K=100$ and $\gamma=1000$.}
\label{tab:score_based_weight_comparison}
\begin{tabular}{lcccccccccccc}
\hline
$m$
& 3 & 5 & 8 & 10 & 20 & 30 & 40 & 50 & 60 & 70 & 80 & 100 \\
\hline
$\log(1+k)$
& 53.97 & 46.10 & 39.87 & 37.23 & 30.36 & 26.68 & 24.48 & 22.59 & 21.33 & 20.19 & 19.33 & 17.75 \\
\hline
$\sqrt{k}$
& 53.34 & 46.28 & 39.95 & 37.54 & 30.60 & 26.93 & 24.44 & 22.65 & 21.18 & 20.11 & 19.07 & 17.44 \\
\hline
$k$
& 53.16 & 44.53 & 38.10 & 35.04 & 26.58 & 22.19 & 19.25 & 17.09 & 15.49 & 14.20 & 13.19 & 11.55 \\
\hline
$k^2$
& 69.67 & 58.54 & 48.17 & 43.58 & 29.01 & 21.88 & 17.71 & 15.00 & 13.04 & 11.64 & 10.49 & 8.83 \\
\hline
$k^3$
& 98.21 & 83.63 & 69.35 & 61.73 & 38.14 & 26.18 & 19.97 & 16.15 & 13.61 & 11.79 & 10.43 & 8.48 \\
\hline
$k^4$
& 129.60 & 110.95 & 92.59 & 84.04 & 49.83 & 31.81 & 22.90 & 17.76 & 14.52 & 12.31 & 10.69 & 8.42 \\
\hline
\end{tabular}
\end{table*}

\begin{table*}[h!]
\centering
\caption{Best-performing rank-weighting functions for score-based statistics with $K=100$ and $\gamma=1000$.}
\begin{tabular}{lcccccccccccc}
\hline
$m$ 
& 3 & 5 & 8 & 10 & 20 & 30 & 40 & 50 & 60 & 70 & 80 & 100 \\
\hline
$k$ 
& 53.16 & 44.53 & 38.10 & 35.04 & 26.58 & 22.19 & 19.25 & 17.09 & 15.49 & 14.20 & 13.19 & 11.55 \\ 
\hline
best delay
& 53.16 & 44.53 & 38.10 & 35.04 & 26.58 & 21.88 & 17.71 & 15.00 & 13.04 & 11.64 & 10.43 & 8.42 \\ 

corresponding weights
& $k$ & $k$ & $k$ & $k$ & $k$ & $k^2$ & $k^2$ & $k^2$ & $k^2$ & $k^2$ & $k^3$ & $k^4$ \\ 
\hline
\end{tabular}
\label{tab:best_performance_score}
\end{table*}
\subsection{GB-RBM Data Generation}

Following \cite{Wuetal_IT_2024,Banerjee_SQA_2024}, we generate the pre-change and post-change observations from Gaussian--Bernoulli Restricted Boltzmann Machines (GB-RBMs). This experimental setting enables us to evaluate the proposed score-based fusion procedure under a non-Gaussian model for which the likelihood ratio is not available in closed form. Such a setting is representative of high-dimensional i.i.d. observations, such as image or feature data acquired from multiple sensors, where explicit probabilistic models are difficult to obtain but score-based methods remain applicable.

Following \cite{Wuetal_IT_2024,Banerjee_SQA_2024}, we adopt the same experimental setting and set the visible and hidden dimensions to
\[
d_x=d_h=5.
\]
The pre-change GB-RBM is parameterized by a weight matrix
\[
W_0\in\mathbb{R}^{5\times5},
\]
and bias vectors
\[
b\in\mathbb{R}^{5},
\qquad
c\in\mathbb{R}^{5},
\]
whose entries are independently generated from the standard normal distribution. Once generated, these parameters remain fixed throughout the experiment.

The post-change distribution is constructed by perturbing only the interaction matrix,
\[
W_1=W_0+\epsilon,
\]
where the entries of $\epsilon$ are independently sampled from $\mathcal{N}(0,0.1^2)$, while the bias vectors $b$ and $c$ remain unchanged.

Samples from both the pre-change and post-change distributions are generated using Gibbs sampling. A total of 30,000 observations are generated from each distribution and are subsequently used for training the score networks described in the next subsection.

\subsection{Score-Based Change Detection}
Using the pre-change and post-change observations generated in the previous subsection, one network is trained using samples from the pre-change distribution, while the other is trained using samples from the post-change distribution \cite{Wuetal_IT_2024}.
Let $P$ and $Q$ denote the pre-change and post-change distributions with probability density functions $p$ and $q$, respectively. The learned score is then substituted into the Hyvärinen score, yielding the statistic
\[
\nabla_x \log p(x), \text{ and } \nabla_x \log q(x).
\]
After training, the network parameters are fixed and are used throughout all change-detection experiments without further updating.

Using the learned score, the corresponding Hyv\"arinen score \cite{hyvarinen2005estimation} is computed as
\[
S_H(X,P)
=
\frac{1}{2}
\left\|
\nabla_x \log p(X)
\right\|^2
+
\Delta_x \log p(X),
\]
where $\Delta_x = \sum_{i=1}^{d_x} \frac{\partial^2}{\partial x_i^2}$
denotes the Laplacian operator with respect to the observation vector $X$. Since the observation dimension is $d_x=5$, the Laplacian term is computed directly.

\vspace{0.5em}
Let
\[
Y(X)
=
S_H(X,Q)-S_H(X,P)
\]
denote the score difference between the post-change and pre-change models. The local score-based statistic is then updated recursively as
\begin{equation}
 R_{k,n}
=
\left(
1+R_{k,n-1}
\right)
\exp \left(
\lambda Y(X_{k,n})
\right),   
\label{eq:score_based_R_k_n}
\end{equation}
where $\lambda > 0 $ is the scaling parameter proposed in \cite{Wuetal_IT_2024}. The local score-based statistics are first ordered according to their magnitudes and then fused using the proposed rank-weighted fusion rule in \eqref{eq:fusion_rule}.

At each time $n$, the local score-based statistics are ordered as
\[
R_{(1),n}
\geq
R_{(2),n}
\geq
\cdots
\geq
R_{(K),n}.
\]
For a fixed rank-weighting function $w(\cdot)$, we calculate the score-based global fusion statistic using \eqref{eq:fusion_rule}.
We employ the same  stopping rule as in \eqref{eq:stopping}, and the threshold $A_w$ is calibrated separately for each weighting function so that the same average run length to false alarm constraint is satisfied,
\[
\mathbb{E}_{\infty}
\left[
T^{(w)}
\right]
=
\gamma.
\]
In our experiments, we use $\gamma=1000$. After this calibration, the detection delays of different weighting functions can be compared under the same false-alarm constraint.

\subsection{Comparison of Rank-Weighting Functions}

We investigate the same family of rank-weighting functions considered in the previous section,
\[
\log(1+k),\quad
\sqrt{k},\quad
k,\quad
k^2,\quad
k^3,\quad
k^4.
\]

For each weighting function, the detection threshold is calibrated separately so that every procedure satisfies the same average run length to false alarm constraint. This separate calibration is necessary because different weighting functions produce score-based fusion statistics on different numerical scales. Consequently, all detection delays are compared under the same false-alarm constraint.

Table~\ref{tab:score_based_weight_comparison} reports the average detection delay for each weighting function as the number of affected streams varies. 
Similar to the SRP-based experiments, the relative performance of the weighting functions depends strongly on the sparsity level of the change. 
When only a small number of streams are affected, the linear weight $w(k)=k$  consistently provides the best or nearly the best performance. 
As the number of affected streams increases, however, higher-order polynomial weights become increasingly effective and yield substantially shorter detection delays.

To better illustrate this trend, Table~\ref{tab:best_performance_score} summarizes, for each value of $m$, the smallest detection delay among all candidate weighting functions together with the corresponding weight. The results show that the linear weight is optimal only when the change is relatively sparse. Specifically, $w(k)=k$ achieves the smallest detection delay for $m=3,5,8,10$, and $20$. When more streams are affected, the quadratic weight $w(k)=k^2$ becomes the best-performing choice for $m=30,\ldots,70$, while $w(k)=k^3$ and $w(k)=k^4$ achieve the smallest delays for $m=80$ and $m=100$, respectively.

Therefore, the score-based experiments lead to the same conclusion as the SRP-based experiments: no single fixed rank-weighting function is uniformly optimal over all sparsity regimes. Although the conventional linear weight performs well when relatively few streams are affected, higher-order polynomial weights provide noticeably shorter detection delays in denser change scenarios. This consistency suggests that the effectiveness of rank-weighted fusion depends primarily on the fusion mechanism rather than on the particular form of the local detection statistic.

\section{Conclusion}
\label{sec:conclusion}
In this paper, we investigated an MoE framework for QCD in multi-stream data. 
By considering different families of mixtures for the RWS scheme, we showed that one can achieve a significant improvement in performance as compared to a fixed mixture of linear weights. 
Numerical experiments based on both local Shiryaev-Roberts statistics and score-based local statistics demonstrated that 
no single fixed rank-weighting function is uniformly optimal across different post-change regimes. 
While the conventional linear weight performs well when relatively few streams are affected, 
higher-order polynomial weights achieve noticeably shorter detection delays when a larger number of streams are affected. 
These observations indicate that the choice of rank-weighting function is an important design component of rank-based fusion and should depend on the underlying change scenario.
These results show the promise of an MoE framework for multi-stream QCD and for QCD in general. 
An important direction for future research is to develop adaptive procedures that automatically select or combine different weighting functions online according to the observed data.


%





\ifCLASSOPTIONcaptionsoff
  \newpage
\fi

\balance

\bibliographystyle{IEEEtran}
\bibliography{IEEEexample}
\end{document}